\documentclass[aps,prl,twocolumn,showpacs,groupedaddress,amsmath,amssymb]{revtex4-1}
\usepackage{graphicx}
\usepackage{subfigure}
\usepackage{color}

\begin{document}

\title{Non-Linear Localized Modes Give Rise to a Reflective Optical Limiter}
\author{Eleana Makri, Hamidreza Ramezani, Tsampikos Kottos}
\address{Department of Physics, Wesleyan University, Middletown CT-06459, USA}
\author{Ilya Vitebskiy}
\address{Air Force Research Laboratory, Sensors Directorate, Wright Patterson AFB, OH 45433 USA}
\date{\today}

\begin{abstract}
Optical limiters are designed to transmit low intensity light, while
blocking the light with excessively high intensity. A typical passive
limiter absorbs excessive electromagnetic energy, which can cause its
overheating and destruction. We propose the concept of a layered reflective
limiter based on resonance transmission via a non-linear localized mode.
Such a limiter does not absorb the high level radiation, but rather reflects
it back to space. Importantly, the total reflection occurs within a broad
frequency range and for an arbitrary direction of incidence. The same
concept can be applied to infrared and microwave frequencies.
\end{abstract}

\maketitle

\address{Department of Physics, Wesleyan University, Middletown CT-06459,
USA}

\address{Air Force Research Laboratory, Sensors Directorate, Wright
Patterson AFB, OH 45433 USA}

%========================================================
The continuing integration of optical devices into modern technology has led to the development of an ever increasing 
number of novel schemes for efficiently manipulating the amplitude, phase, polarization, or direction of optical beams 
\cite{ST91}. Among these manipulations, the ability to control the intensity of light in a predetermined manner is of the 
utmost importance, with applications ranging from optical communications to optical computing \cite{O97,PHR98} and 
sensoring. As laser technology is making progress, novel protection devices (optical limiters) are needed to protect
optical sensors and other components from high-power laser damage \cite{limiter1,limiter2,limiter3}. 

Here we focus on the most popular, passive optical limiters. The simplest realization of a passive optical limiter 
is provided by a single nonlinear layer with the imaginary part $n^{\prime \prime }$ of the refractive index being 
dependent on the light intensity $W$. At low intensity, the value $n^{\prime \prime }(W)$ is relatively small, and 
the nonlinear layer is transparent. As the light intensity exceeds certain level, the value $n^{\prime \prime }(W)$ 
increases dramatically, and the nonlinear protective layer turns opaque. In more sophisticated schemes, the nonlinear 
layer can be a part of a complicated optical setup. The problem though is that in all cases, the limiter absorbs the 
excessive power, which might cause overheating or even destruction of the device (a sacrificial limiter). Our goal 
is, using the existing nonlinear materials, to design a photonic structure which would reflect the excessive power 
back to space, rather than absorbing it. Such a structure can be referred to as a passive reflective limiter. A free-
space realization of a reflective limiter is a layered array reflecting a high intensity light regardless of the direction 
of incidence and within a broad frequency range.

\bigskip 
%----------------------------------------------------------------------------------------------
\begin{figure}[th]
\includegraphics[width=1\linewidth, angle=0]{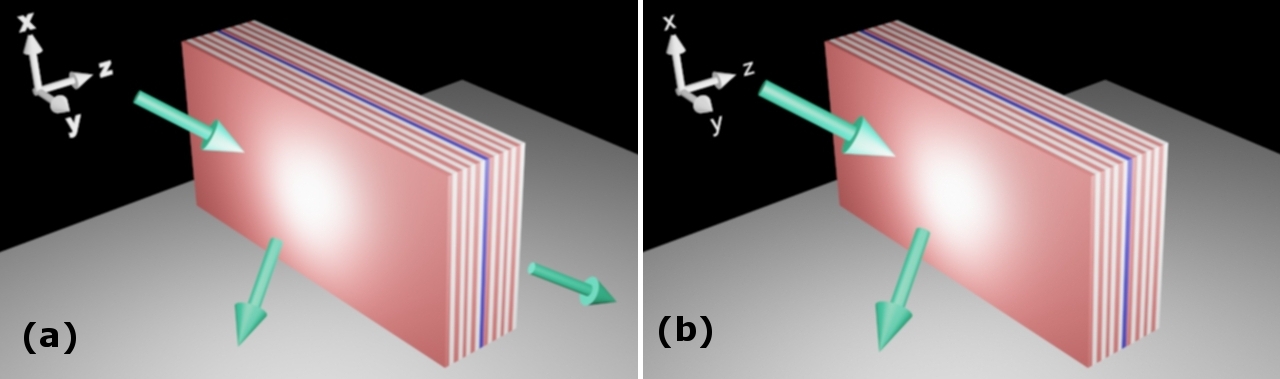}
\caption{(Color online) A power limiter consisting of a non-linear lossy
layer (blue layer) embedded in a Bragg grating (white and orange layers).
This set-up allows for (a) a transmission of a low intensity beam while (b)
it completely reflects a high intensity beam without any absorption. }
\label{fig1}
\end{figure}
%----------------------------------------------------------------------------------------------

Our proposal is based on resonance transmission through a nonlinear localized mode. The simplest realization of the above idea is illustrated in
Fig. \ref{fig1}, where a nonlinear lossy layer is sandwiched between two linear lossless Bragg reflectors. We will show that if the light intensity
is low, the absorption can be small, and the layered structure in Fig. \ref{fig1} will be transmissive in the vicinity of the localized mode frequency.
If the incident light intensity exceeds a certain level, the non-linear layer in Fig. \ref{fig1} decouples the two Bragg gratings and the entire
stack becomes highly reflective -- not opaque, as in the case of a stand-alone nonlinear layer. In other words, the high intensity light will
be reflected back to space, rather than absorbed by the limiter. Even this simple design provides a broad band protection for an arbitrary direction
of incidence. For a given nonlinear material, the intensity limitation of the transmitted light can be controlled by adjusting the layered structure,
so that the electromagnetic energy density in the vicinity of the nonlinear layer is either enhanced or attenuated. A problem with the simple design
of Fig. \ref{fig1} is that the low-intensity transmittance only occurs in the vicinity of the localized mode frequency. This problem can be addressed
by using more sophisticated photonic structures, for instance, those involving two or more coupled defect layers, as it is done in the case of
optical filters \cite{M01}.

%----------------------------------------------------------------------------------------------------
To illustrate our idea, we consider a pair of identical Bragg gratings, each consisting of two alternating layers with real permittivities $\epsilon _{1}$
and $\epsilon _{2}$, placed in the intervals $-L\leq z\leq 0$ and $d_{\gamma}\leq z\leq L+d_{\gamma }$. The width of each grating layer is $d$. A
non-linear lossy layer of width $d_{\gamma }$ is placed between the two gratings at $0\leq z\leq d_{\gamma }$; its complex permittivity $\epsilon
_{\gamma }=\epsilon (1+i\gamma |E(z)|^{2})$ is field dependent. In the particular case of $\epsilon =\epsilon _{1}$ and $\gamma =0$, we have a
standard Bragg grating with a band-gap around the frequency $\omega_{B}=c/(n_{0}d)$ ($c$ is the speed of light). The defect layer creates a localized 
mode with the frequency 
$\omega _{r}$ lying within a photonic band-gap. At this frequency, the entire stack displays resonance transmission accompanied by a dramatic
field enhancement in the vicinity of the defect layer. The enhanced field, in turn, causes the respective increase in the imaginary part of the defect
layer permittivity, $\epsilon _{\gamma }$. The latter will eventually result in decoupling of the two Bragg reflectors and rendering the entire structure
in Fig. \ref{fig1} highly reflective.

We first consider normal incidence. In this arrangement, a time-harmonic electric field of frequency $\omega$ obeys the Helmholtz equation: 
\begin{equation}  \label{Helmholtz}
{\frac{\partial^2 E (z)}{\partial z^2}} + {\frac{\omega^2 }{c^2}}
\epsilon(z) E(z) = 0\,\,\,.
\end{equation}
Eq.~(\ref{Helmholtz}) admits the solution $E_0^{-}(z)=E_{f}^- \exp(ikz)+E_{b}^- \exp(-ikz)$ for $z<-L$ and $E_0^{+} (z)=E_{f}^+ \exp(ikz) + E_{b}^+
\exp(-ikz)$ for $z>L+d_{\gamma}$ where the wavevector $k= n_0\omega/c$. The transmittance, reflectance and absorption, say for a left incident 
wave, are then defined as $\mathcal{T}=|E_f^+/E_f^-|^2$; $\mathcal{R}=|E_b^- /E_f^-|^2$; and $\mathcal{A}=1-\mathcal{T}-\mathcal{R}$ respectively 
\cite{note1}. They can be calculated numerically using a backward map approach.

%---------------------------------------------------------------------------------------------
\begin{figure}[tbp]
\begin{center}
\includegraphics[width=1\linewidth, angle=0]{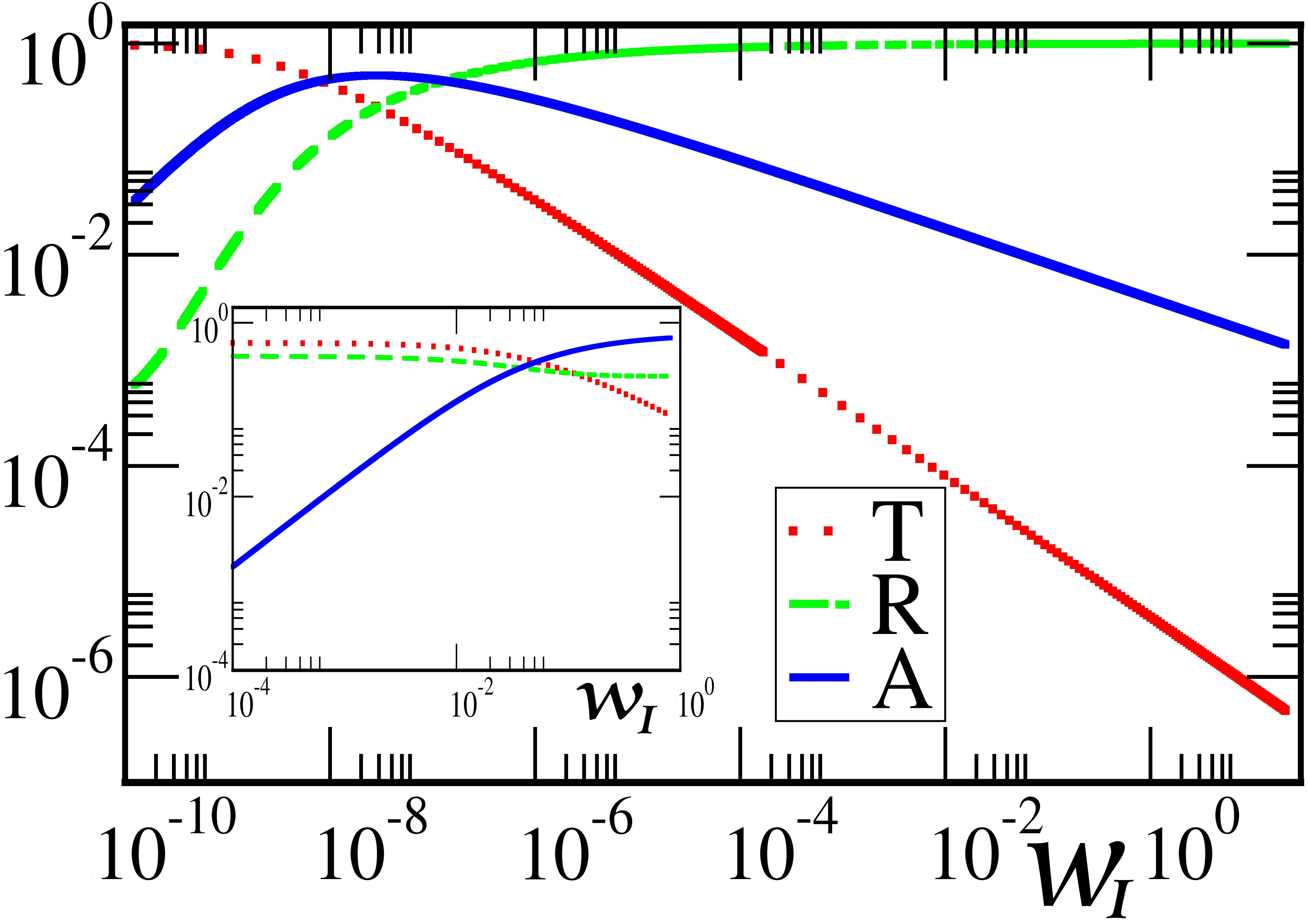}
\end{center}
\caption{Normal incidence for the structure of Fig. \protect\ref{fig1}. We report the transmittance $\mathcal{T}$, 
absorption $\mathcal{A}$ and reflectance $\mathcal{R}$, as a function of the incident power $\mathcal{W}_I $ at 
a resonant frequency $\protect\omega_r\approx 8.15$. The parameters of the grating are indicated at the text. We 
observe that for moderate values of $\mathcal{W}_I$, both $\mathcal{T}$ and $\mathcal{A}$ are suppressed and 
the system becomes reflective i.e. $R\approx1$. Inset: $\mathcal{T}$, $\mathcal{A}$, $\mathcal{R}$ for a single 
non-linear layer (normal incidence). This system, for moderate $\mathcal{W}_I$-values, does not reflect but mainly 
absorbs the incident energy.}
\label{fig2}
\end{figure}
%----------------------------------------------------------------------------------------------------

The amplitudes of forward and backward propagating waves on the left $z<-L$
(right $z>L+d_{\gamma}$) domains outside of the grating are related to the
ones before (after) the non-linear impurity layer by the algebraic
relations: 
\begin{equation}
\left(
\begin{array}{c}
E_{f}^b \\ 
E_{b}^b
\end{array}
\right)= M^{(L)} \left(
\begin{array}{c}
E_{f}^- \\ 
E_{b}^-
\end{array}
\right);\,\, \left(
\begin{array}{c}
E_{f}^+ \\ 
E_{b}^+
\end{array}
\right)= M^{(R)} \left(
\begin{array}{c}
E_{f}^a \\ 
E_{b}^a
\end{array}
\right)  \label{transfer}
\end{equation}
where $M^{(L)}$ ($M^{(R)}$) are the $2\times 2$ transfer matrices of the optical structure associated with the domain 
$-L\leq z \leq0$ ($d_{\gamma}\leq z\leq L+d_{\gamma}$). Above we have expressed the field before (after) the nonlinear 
layer as $E^{b}=E_{f}^b \exp(ikz)+E_{b}^b \exp(-ikz)$ ($E^{a}=E_{f}^a \exp(ikz) + E_{b}^a \exp(-ikz)$). The field $E^{a}(z
=d_{\gamma})$ and its derivative $(dE^{a}/dz)|_{z=d_{\gamma}}$ just after the non-linear layer is then evaluated using 
$M^{(R)}$ from Eq.~(\ref{transfer}) together with the boundary conditions (associated with a left incident wave) $E_b^+ 
= 0$ and $E_f^+=1$. Using $E^{a}(z=d_{\gamma})$ and $(dE^{a}(z)/dz)|_{z=d_{\gamma}}$ as initial conditions we have 
integrated backwards Eq. (\ref{Helmholtz}), with the help of a 4rth order Runge-Kutta, and obtained the field $E^{b}(z=0)$ 
and its derivative $(dE^{b}/dz)|_{z=0}$ at the other end $z=0$ of the nonlinear layer. From these values we evaluate the 
forward $E_{f}^b$ and backward $E_{b}^b$ propagating amplitudes. Utilizing Eq. (\ref{transfer}) together with $M^{(L)}$ 
we finally find the amplitudes $E_{f}^-$ and $E_{b}^-$ which allow us to calculate $\mathcal{T}, \mathcal{R}$ and 
$\mathcal{A}$. Note that for a backward map with boundary condition $E_f^+=1$ we have $|E_f^-|^2=1/\mathcal{T}$.

It is convenient to work with the rescaled variable ${\tilde E}(z)=\sqrt{\gamma} E$. In this representation, Eq. (\ref{Helmholtz}) 
becomes 
\begin{equation}  
\label{Helmholtz2}
{\frac{\partial^2 {\tilde E} (z)}{\partial z^2}} + {\frac{\omega^2 }{c^2}} {
\tilde \epsilon}(z) {\tilde E}(z) = 0\,\,\,
\end{equation}
where ${\tilde \epsilon}(z\notin [0,d_{\gamma}])=\epsilon(z)$, while ${\tilde \epsilon}(z\in[0,d_{\gamma}])= \epsilon_{\gamma} 
(1+i|{\tilde E}(z)|^2)$. In other words, in this representation, the non-linear layer has a fixed absorption rate which is equal 
to unity, the outgoing field boundary associated with the backward map varies as ${\tilde E}_f^+=\sqrt{\gamma}$ while 
the incident light intensity $\mathcal{W}_I$ is $\mathcal{W}_I\equiv |{\tilde E}_f^-|^2=\gamma/\mathcal{T}=\gamma |E_f^-|^2$.

In Fig. \ref{fig2}, the effect of the incident intensity $\mathcal{W}_I$ on the transmission, reflection and absorption of a 
resonant localized mode is presented. The Bragg grating used in these simulations consists of 40 layers on each side with 
alternating permittivities $\epsilon_1=4$ and $\epsilon_2=9 $. The width of the impurity layer is $d_{\gamma}=1$ and the
amplitude $\epsilon$ of the nonlinear permittivity is $\epsilon=9$. We have confirmed numerically that in the linear case 
the defect creates a resonant mode at $\omega_r\approx 8.15$ \cite{note4} at the band-gap of the grating which is
localized around the impurity. We find that as the incident intensity $\mathcal{W}_I$ increases (main panel of Fig. \ref{fig2}), 
the transmittance of this resonant mode decreases, with a simultaneous increase of the absorption. Further increase of 
$\mathcal{W}_I$, results in noticeable growth of the reflectance with a simultaneous decrease of the absorption and
transmittance. Eventually both $\mathcal{T}$ and $\mathcal{A}$ vanishes for moderate values of $\mathcal{W}_I$. In 
other words the system reflects completely the incident radiation. For the shake of comparison we also calculated 
$\mathcal{T}, \mathcal{A}$ and $\mathcal{R}$ versus $\mathcal{W}_I $ for a single non-linear layer with no Bragg 
reflectors (see inset of Fig. \ref{fig2}). We find that for the same range of moderate values of incident intensity 
$\mathcal{W}_I$, the system rather absorbs the energy instead of reflecting it back to space.

%----------------------------------------------------------------------------------------------------
For normal incidence, a further theoretical analysis can be carried out. To this end we assume that the permittivity of the 
non-linear layer is $\epsilon_{\gamma}(z) = \epsilon (1 + i\gamma |E(z)|^2)\delta(z)$. This approximation is justified in 
the case of a \textit{thin} metallic defect. For the analytical calculation of $\mathcal{T}, \mathcal{R}$ and $\mathcal{A}
$, we proceed along the same lines that we have highlighted in the numerical analysis previously. For the sake of generality 
we will assume that the transport characteristics of the left and right linear subsystems are encoded in the values of their 
left (right) transmission $t_L (t_R)$ and reflection $r_L (r_R)$ amplitudes. The elements of the transfer matrices $M^{(L)}$ 
and $M^{(R)}$ (see Eq. (\ref{transfer})) are defined as $M_{11}^{(L/R)}=1/t_{L/R}^*$, $M_{12}^{(L/R)}=-r_{L/R}/t_{L/R}$, 
$M_{21}^{(L/R)}=-(r_{L/R}/t_{L/R})^*$, and $M_{22}^{(L/R)}=1/t_{L/R}$.

Next we calculate the field amplitudes just before and after the delta defect by utilizing the transfer matrices Eq.(\ref{transfer}) 
associated with the linear segments. For a left incident wave, we have at $z=0^-$ 
\begin{equation}  \label{eq1}
E_f^b =\frac{E_f^-}{t_L^{*}}-\frac{r_L E_f^+}{t_L} ;\quad E_b^b =\frac{E_f^+}{t_L}-\frac{E_f^-r_L^{*}}{t_L^{*}}
\end{equation}
while at $z=0^+$ just after the delta defect we have 
\begin{equation}  \label{eq2}
E_f^a =\frac{t_R^{*} E_f^+}{1-|r_R|^2} ;\quad E_b^a =\frac{t_R r_R^{*} E_f^+}{1-|r_R|^2}.
\end{equation}
Using Eqs. (\ref{eq1},\ref{eq2}) together with the continuity of the field at $z=0$ and the suitable discontinuity of its derivative 
we write the incident and reflected field amplitudes in terms of the transmitted wave amplitude 
\begin{equation}
\begin{array}{ccc}
E_f^- & = & \{\frac{1}{\tau_0}-i(\frac{1}{\tau}-\frac{1}{\tau_0})\gamma
|\xi|^2|E_f^+|^2\}E_f^+ \\ 
E_b^- & = & (\frac{t_L}{1-r_L})\{ \xi E_f^+-\frac{(1-r_L ^{*})}{t_L ^{*}}
E_f^-\}. \\ 
&  & 
\end{array}
\label{non}
\end{equation}
Above, $\tau$ is the transmission amplitude in the absence of the $\delta-$like layer, $\tau_0$ is the transmission amplitude 
when $\gamma=0$, and $\xi=\frac{t_R^{*}+t_Rr_R^{*}}{1-|r_R|^2}$. From Eq.(\ref{non}) we deduce the transmission, reflection 
and absorption amplitudes. For the transmission and reflection amplitude we get that 
\begin{equation}
t=\frac{1}{\frac{1}{\tau_0}-i(\frac{1}{\tau}-\frac{1}{\tau_0})\gamma |\xi|^2
|E_f^+|^2}; r=(\frac{t_L}{1-r_L})\{ t\xi-\frac{1}{t_L ^{*}}(1-r_L ^{*})\}.
\label{trans}
\end{equation}
The transmittance, reflectance and absorption can then be calculated as $\mathcal{T}=|t|^2, \mathcal{R}=|r|^2$ and $\mathcal{A}
=1- \mathcal{T}-\mathcal{R}$. From Eq. (\ref{trans}) we observe that increasing $\gamma$ (we remind that the incident light intensity 
$\mathcal{W}_I\sim \gamma$) results in an increase of the denominator of the transmission amplitude and therefore to a decrease 
of $\mathcal{T}$ (for very large $\gamma$-values it becomes zero). At the same time the reflection amplitude, becomes $r\rightarrow 
(\frac{t_L}{1-r_L})\{-\frac{1}{t_L ^{*}}(1-r_L ^{*})\}$ corresponding to perfect reflection, i.e. $\mathcal{R} \rightarrow 1$. Consequently 
in this limit we have zero absorption $\mathcal{A}=0$.

%----------------------------------------------------------------------------------------------
\begin{figure}[tbp]
\begin{center}
\includegraphics[width=1\linewidth, angle=0]{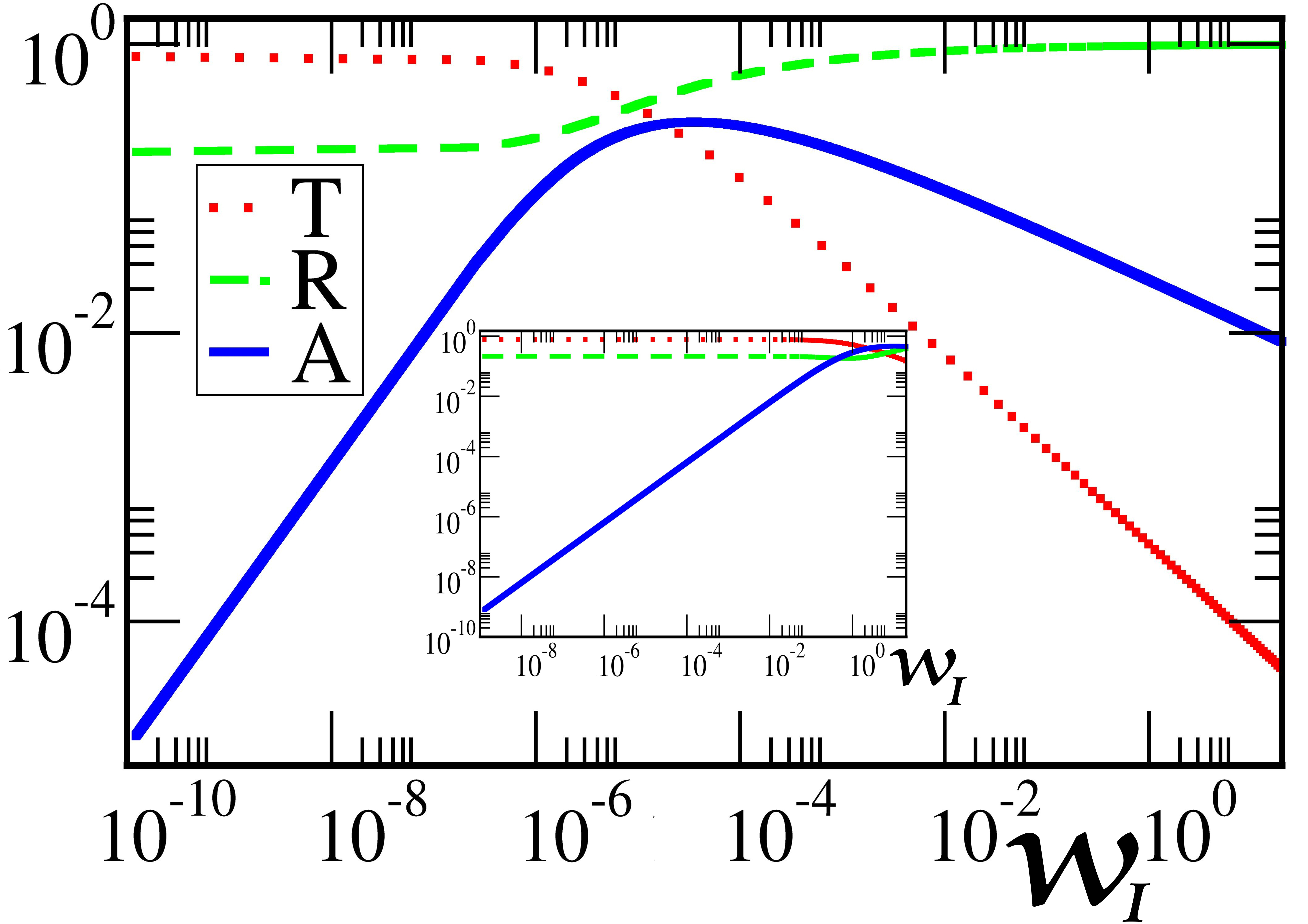}
\end{center}
\caption{(Color online) Transport characteristics for the model of a
non-linear $\protect\delta-$like defect embedded in a Bragg grating. The
theoretical results of Eq. (\protect\ref{trans}) shown here, reproduce
nicely the features of the simulations reported in Fig \protect\ref{fig2}.
In the inset we report, for comparison, $\mathcal{A},\mathcal{T}$ and $\mathcal{R}$ for a single non-linear layer. We assume normal 
incidence at $\protect\omega=0.7\approx\protect\omega_r$. }
\label{fig3}
\end{figure}
%------------------------------------------------------------------------------------------------

Fig.~\ref{fig3} demonstrates the effect of $\mathcal{W}_I$ on a resonant localized mode for the case of symmetrically placed Bragg 
gratings on the left and right side of a $\delta-$like defect. The alternate layers at the Bragg gratings have permittivity $\epsilon_1=4$ 
and $\epsilon_2=9$ while the permittivity of the defect layer is $\epsilon=1.5$. The transport characteristics of the gratings $t_L=t_R$ 
and $r_L=r_R$ have been calculated numerically and used as inputs in Eqs. (\ref{trans}). We find (see Fig. \ref{fig3}) that the overall 
behavior of $\mathcal{T}$, $\mathcal{R}$ and $\mathcal{A}$ is similar to the one observed in the simulations of Fig. \ref{fig2}.

For comparison, we also report (inset of Fig. \ref{fig3}) the behavior of $\mathcal{T}, \mathcal{A}$ and $\mathcal{R}$, for a single 
non-linear layer (without any Bragg gratings), vs. the incident intensity $\mathcal{W}_I$. They are calculated analytically using the 
continuity of the field and the discontinuity of its derivative at the position of the $\delta-$defect. Specifically, $\mathcal{T} = 
\frac{4}{(k \epsilon_0)^2+(2+k \epsilon_0\gamma |E_f^+|^2 )^2};\, \mathcal{R} = (k \epsilon_0)^2 (1+\gamma^2 |E_f^+|^4)
\mathcal{T}/4$ and $A =k \epsilon_0 \gamma |E_f^+|^2 \mathcal{T} $. We find that for moderate $\mathcal{W}_I$-values the 
single non-linear layer is mainly absorptive (inset of Fig. \ref{fig3}) while the structure of Fig. \ref{fig1} is mainly reflecting the 
incident light back to space (main panel of Fig. \ref{fig3}).

%---------------------------------------------------------------------------------------------------------------
\begin{figure}[tbp]
\includegraphics[width=.85\linewidth, angle=0]{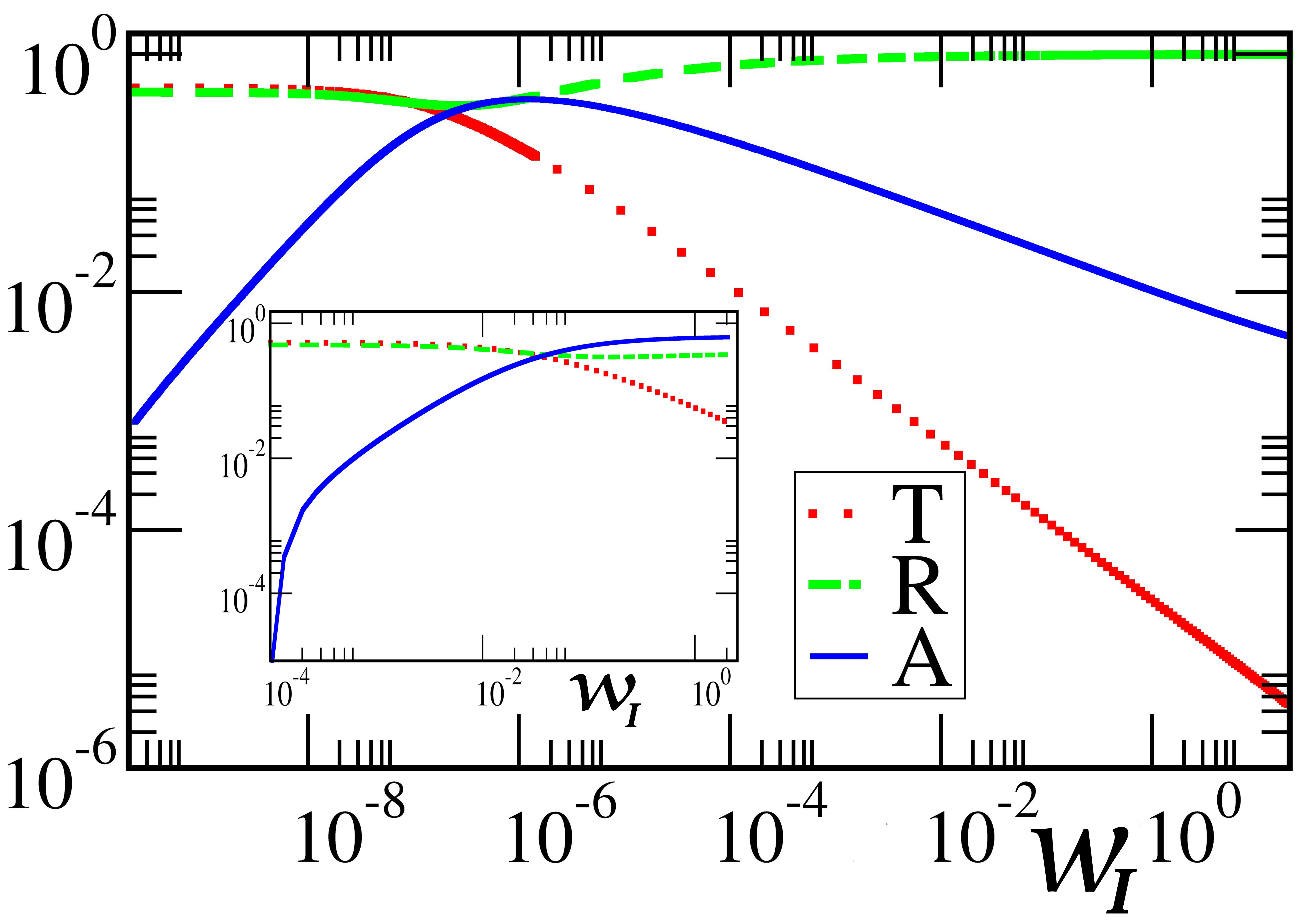}
\caption{(Color online) Simulations for the structure of Fig. \protect\ref{fig1} for oblique incidence at $\protect\omega=8.95
\approx \protect\omega_r$. The parameters of the grating are indicated at the text while the incident angle is $\protect\phi
=6^o$. We find that for moderate values of the incident light intensity $\mathcal{W}_I$, the transmittance and absorption
are suppressed and the system is reflective i.e. $\mathcal{R}=1$. In the inset we report for comparison (and for the same range 
of $\mathcal{W}_I$-values) the $\mathcal{A}, \mathcal{T}, \mathcal{R}$ values for the case of a single non-linear layer. This 
system mainly absorbs the incident energy. }
\label{fig4}
\end{figure}
%---------------------------------------------------------------------------------------------------------------

%----------------------------------------------------------------------------------------------------
We have also investigated the efficiency of the proposed limiter in the case of oblique incidence \cite{MRVK13}. A representative 
example in the case of an incident angle $\phi=6^o$ is shown in Fig. \ref{fig4}. The Bragg grating considered in this example 
consists of two layers with permittivities $\epsilon_1=9$, $\epsilon_2=16$ while the non-linear impurity has permittivity 
$\epsilon=16$. We find again that as the incident light intensity $\mathcal{W}_I$ takes moderate values, the transmittance and 
the absorption are suppressed and the structure becomes reflective i.e. $\mathcal{R}\approx 1$. This behavior has to be contrasted 
with the one found for the single non-linear layer where for moderate $\mathcal{W}_I$-values the dominant mechanism is absorption, 
see the inset of Fig. \ref{fig4}.

%----------------------------------------------------------------------------------------------------
The effectiveness of the structure of Fig.~\ref{fig1} to act as a self-protecting power limiter for any incident angle calls for a generic
argument for its explanation. The following heuristic argument, provides some understanding of the mechanism underlying our 
structure. First we recall that the defect results in the creation of a resonance mode which is localized around the impurity layer 
at $z=0$ and decays away from its localization center with an envelope profile $E_{\mathrm{r}}(z)\sim\exp(-\alpha |z|)$. An 
incoming (say from the left) wave that carries an incident energy flux $\mathcal{S}$ can resonate via this mode as long as the
loss coefficient is $\gamma\leq \gamma^*\sim \mathcal{S}/{\cal W}_0 \sim \exp(-2\alpha L)$ (${\cal W}_0\sim |E_{\mathrm{r}}
(z=0)|^2\sim\exp(2\alpha L) $ is the mode intensity at $z=0$ \cite{note2}). In other words the energy that is absorbed from the 
non-linear lossy layer via the resonant mode cannot be more than the incoming energy. Therefore for any $\gamma>\gamma^*$ 
the resonant mode will not be sustained and thus the transmission will be $\mathcal{T}=0$.

We proceed in our argument by noticing that the resonant mode is located at the band-gap of the Bragg grating and therefore it 
can be written as a superposition of two evanescent modes, one growing and another one decaying i.e. $E_{\mathrm{r}}(z)\sim 
\psi_+(z) + \psi_-(z)$, where $\psi_- \sim \alpha_-\exp(-z)$ and $\psi_+\sim \alpha_+\exp(z)$. Let us assume that $\alpha_+
\sim \mathcal{O}(1)$ \cite{note3}. Then the field at the outer boundary of the left grating at $z=-L$ is $E_{\mathrm{r}}( z=-L)=
\alpha_{+}\exp(-L)+\alpha_{-} \exp(L)\sim \alpha_{-}\exp(L)$. At the same time due to continuity at the boundary we expect that 
the resonance wavefunction must be equal to the incoming field which we assume to take some constant value i.e. $\alpha_{-}
\exp(L) \sim \mathcal{O} (1)$. This can only happen if $\alpha_{-} \rightarrow 0$. Finally we recall that the incoming energy flux 
is given by the Poynting vector $\mathcal{S}$ which in the case of evanescent modes is $S\sim\psi_{+}\psi_{-}=\alpha_{+}
\alpha_{-}\rightarrow 0$ \cite{Ilya}. Therefore there will be no net flux towards the structure and thus $\mathcal{A}=0$.
Since $\mathcal{T}=0$ and $\mathcal{A}=0$ we conclude that almost all the incident energy is reflected back i.e. $\mathcal{R}\rightarrow 1$.

%----------------------------------------------------------------------------------------------------
In conclusion, we have examined the scattering problem for a periodic
layered structure with an embedded nonlinear defect layer. We presume that
the absorption coefficient of the defect layer increases with the light
intensity, which is normally the case. We have shown that such a layered
structure acts as a self-protecting power limiter. Specifically, at low
intensity of the incident light, the entire stack is highly transmissive.
When the light intensity increases, the stack transmission decreases.
Initially, the fraction of the input power absorbed by the lossy nonlinear
layer also increases with the incident light intensity. But when the input
power exceeds a certain level, the stack becomes highly reflective within a
broad frequency range and regardless of the angle of incidence. In other
words, the excessive radiation will be reflected back to space, rather than
being absorbed by the limiter, which can prevent overheating and destruction
of the limiter. A simplest realization of such a self-protected (reflective)
power limiter is provided by a lossy non-linear layer sandwiched between two
Bragg gratings. A shortcoming of such a design is that although the high
intensity radiation will be reflected back to space within a broad frequency
range, the low-intensity transmittance only occurs within a narrow frequency
band corresponding to the frequency of the localized mode. This problem can
be addressed by using a more sophisticated, structured defect layer, as well
as a chain of several coupled nonlinear defects. The latter possibilities
are currently under investigation \cite{MRVK13}.

%----------------------------------------------------------------------------------------------
\textit{Acknowledgments -} This work is sponsored by the Air Force Research
Laboratory (AFRL/RYDP) through the AMMTIAC contract with Alion Science and
Technology, and by the Air Force Office of Scientific Research LRIR09RY04COR
and FA 9550-10-1-0433.

%--------------------------------------------------------------------


\begin{thebibliography}{99}
\bibitem{note1} The quantities $T,R,\mathcal{A}$ are the same for a right
incident wave as well, since our structure is reciprocal.
\bibitem{note4} In all numerical simulations in Figs. \ref{fig2}-\ref{fig4} we use units such that $c=1;d=1$.
\bibitem{note2} We assume the incident wave has amplitude $\mathcal{O}(1)$
and that due to continuity of the wavefunction at the boundary $E_{\mathrm{%
res}}(z=-L)\sim \mathcal{O}(1)\rightarrow E_{\mathrm{res}}(z=0)\sim \exp(L)$
\bibitem{note3} A similar argument can be used for $\alpha_{-}\sim \mathcal{O%
}(1)$.
\end{thebibliography}
\end{document}